\documentclass[10pt]{article}
\usepackage[letterpaper]{geometry}
\usepackage{hicss}
\usepackage{times}
\usepackage[none]{hyphenat}
\usepackage{url}
\usepackage{latexsym}
\usepackage{indentfirst}
\usepackage{graphicx}
\graphicspath{{images/}}
\usepackage[caption=false]{subfig}
\usepackage[style=apa]{biblatex}
\usepackage{listings}
\usepackage{pifont} 
\usepackage{colortbl}

\usepackage{xcolor}
\colorlet{punct}{red!60!black}
\definecolor{background}{HTML}{EEEEEE}
\definecolor{delim}{RGB}{20,105,176}
\colorlet{numb}{magenta!60!black}

\lstdefinelanguage{json}{
    basicstyle=\small\ttfamily,
    breaklines=true,
    literate=
     *{0}{{{\color{numb}0}}}{1}
      {1}{{{\color{numb}1}}}{1}
      {2}{{{\color{numb}2}}}{1}
      {3}{{{\color{numb}3}}}{1}
      {4}{{{\color{numb}4}}}{1}
      {5}{{{\color{numb}5}}}{1}
      {6}{{{\color{numb}6}}}{1}
      {7}{{{\color{numb}7}}}{1}
      {8}{{{\color{numb}8}}}{1}
      {9}{{{\color{numb}9}}}{1}
      {:}{{{\color{punct}{:}}}}{1}
      {,}{{{\color{punct}{,}}}}{1}
      {\{}{{{\color{delim}{\{}}}}{1}
      {\}}{{{\color{delim}{\}}}}}{1}
      {[}{{{\color{delim}{[}}}}{1}
      {]}{{{\color{delim}{]}}}}{1},
}

\title{Codeless App Development: Evaluating A Cloud-Native Domain-Specific Functions Approach}

 \author{Chuhao Wu \\
  The Pennsylvania State University \\
  {\underline{cjw6297@psu.edu}} \\ \\
  Adrian Mos \\
  Naver Labs Europe \\
  {\underline{adrian.mos@naverlabs.com} } \\ \And
  Jose Miguel Perez-Alvarez \\
  Naver Labs Europe \\
  {\underline{jm.perez@naverlabs.com} } \\ \\
  John M. Carroll\\
  The Pennsylvania State University \\
  {\underline{jmc56@psu.edu} } \\}

\date{June 2022}

\begin{document}
\maketitle

\begin{abstract}

Mobile applications play an important role in the economy today and there is an increasing trend for app enablement on multiple platforms. However, creating, distributing, and maintaining an application remain expert tasks. Even for software developers, the process can be error-prone and resource-consuming, especially when targeting different platforms simultaneously. Researchers have proposed several frameworks to facilitate cross-platform app development, but little attention has been paid to non-technical users. In this paper, we described the Flow framework, which takes the advantage of domain-specific languages to enable no-code specification for app modeling. The cloud-native coordination mechanism further supports non-technical users to execute, monitor, and maintain apps for any target platforms. User evaluations were conducted to assess the usability and user experience with the system. The results indicated that users can develop apps in Flow with ease, but the prototype could be optimized to reduce learning time and workload.

\end{abstract}

\subsubsection*{Keywords:}

Cross-Platform App Development,  Domain Specific Language, Model-Driven Software Development, Cloud Execution, User Experience

\section{Introduction}
Since the release of Apple's first iPhone in 2007, the popularity of mobile applications (apps) has grown rapidly in the last decades. By 2020, there are more than 8.9 million apps available around the world (\cite{koetsierThereAreNow2020}). Global revenue from mobile applications increased to more than 318 billion US dollars in 2021 and is estimated to reach around 613 billion by 2025. Therefore, the presence in app marketplaces is critical for many individuals in today's economy. 

Nowadays, there are various tools to assist non-technical people (i.e. people without specific technical knowledge) to create websites, yet most of them only enable static content and app development still remains a task primarily for programmers. This high threshold of app development could hinder non-technical people from growing their business and suppress their voices in the design and development of apps. The expensive and time-consuming app development process can make it difficult for small businesses to take advantage of the Internet. Therefore, lowering the barriers for non-technical people will enable underrepresented population to participate more actively in the app economy and shape the digital future.  

Apps on traditional mobile devices (i.e. smartphones and tablets) are mostly developed for Android and iOS. However, the electronic devices owned by users have been increasing, and there is a trend towards app-enablement in various devices, including smart TVs, wearables, voice assistants, etc (\cite{rieger2017taxonomy}). Native development approach using tools and languages designed for a specific platform will require extensive cost to deploy an app on multiple platforms. Therefore, cross-platform solutions to simplify app development have attracted significant attention from practitioners and academia in the past decades (\cite{rieger2019towards}). Yet empirical verification is demanded for evaluating cross-platform frameworks, with a particular focus on qualitative user-oriented research (\cite{biorn2018survey}).

In this study, we addressed the research gaps by proposing and evaluating a no-code app development system: \emph{Flow}, which allows non-technical users to create interactive apps. In this system, the non-technical user models or adapts the logic of an app, by using high-level behavioral models.
The novelty of this system resides in that Flow does not transform these models into executable code for target platforms. Instead, Flow executes those models directly on the cloud, and uses a coordination mechanism with the final app when user-iteration is required. This allows the creation of generic applications that are can be easily adapted to different platforms. In this way, Flow makes it easier for people without technical skills to develop and maintain applications across platforms. Users do not need to understand the cloud-based logic and can simultaneously see the effects of changes they make to the models.

The paper is organized as follows: Section 2 presents related work in cross-platform app development. Section 3 introduces the overall architecture and definitions of Flow. Section 4 discusses the setup and results of user evaluation. In Section 5, the implications of Flow are discussed further before concluding with a summary in Section 6.

\section{Related Work}
In general, previous research has distinguished five approaches for developing cross-platform applications (\cite{biorn2018survey}). First, the progressive web approach essentially develops a web application optimized for the mobile device screen. The app cannot be installed on the mobile device and needs to be executed within a browser app. Progressive Web Apps (PWA) introduced by Google integrates the services workers, a manifest file, and a bundle of static user interface components not dependent on dynamic content, which results in a performance comparable to that of native apps (\cite{biorn2017progressive}). However, even with enhanced capabilities, the access to device or platform features are still limited for web apps. 

The hybrid approach allows developers to develop mobile apps using web development tools such as HTML, CSS, and JavaScript. Apache Cordova (\cite{bosnic2016development}) is one of the most popular tools for developing hybrid apps. It initializes the Native app with a WebView (embeddable web browser) and the communication between the WebView and Native code, enabling the specification of business logic through JavaScript and interfaces through HTML and CSS. In addition, libraries such as Ionic Framework (\cite{yusuf2016ionic}) and Sencha Touch (\cite{clark2013sencha}) facilitate the development of user interfaces (UIs) for hybrid apps and make them adhering to the guidelines of the targeted platforms. 

By contrast, the interpreted approach relies on a runtime component, and app developers can access the underlying functionality through the API (\cite{biorn2020empirical}). The app is typically written using programming languages such as JavaScript and the invocation of foreign function interfaces is achieved through proprietary plugin-based bringing systems (e.g., React Native and NativeScript (\cite{brito2018javascript})). Consequently, plugin developed for one system may not work in the other, which fragments the frameworks and tools and becomes a major drawback for this approach. 

The cross-compiled approach maps the input application to a target platforms through compiling to Native byte code. The access to Native device features is exposed to app developers through the Software Development Kit
(SDK) rather than the bridging components in hybrid and interpreted approaches. The generated user interfaces are rendered as Native interface components (\cite{willocx2015quantitative}). One of the most recent cross-compile frameworks is Google's Flutter (\cite{windmill2020flutter}), which recreates the appearance of user interfaces through Skia Canvas. 

The model-driven approach uses the high level abstract representation to develop apps. It utilizes textual or graphical Domain Specific Languages (DSLs) or the Unified Modeling Language (UML) to construct models. Code generators convert the models into Native source code for the targeted platforms. Ideally, the generated apps will be the same as native apps. Numerous model-driven frameworks exist in both industry and academia. For example, the MD\textsuperscript{2} (\cite{heitkotter2013cross}) framework uses textual DSL to specify the components of an app and does not require any knowledge of platform-specific programming languages. However, most model-driven frameworks develop distinct DSLs, which makes it difficult to transfer the knowledge from one to another (\cite{le2013yet}).

Despite the advantages of aforementioned approaches, most of them are not friendly to non-technical people. The model-driven approach may be appropriate for non-technical people, as it does not require a background in any programming language. Moreover, there is a lack of usability evaluation in cross-platform app development research. \textcite{rieger2018evaluating} proposed the Münster App Modeling Language (MAML) framework and evaluated it with software developers, process modelers, and domain experts, resulting in better comprehensibility and usability than other frameworks. Based on the advantage of DSLs, our work further improves the experience for non-technical users by introducing the mechanism to isolate client-side models from changes on the server-side, which makes it easier to develop, deploy, and maintain apps. We conducted a rigorous user evaluation to identify the level of usability and possible improvements from non-technical users' perspective. 

\begin{figure*}[h]
	\centering
	\includegraphics[width=\textwidth]{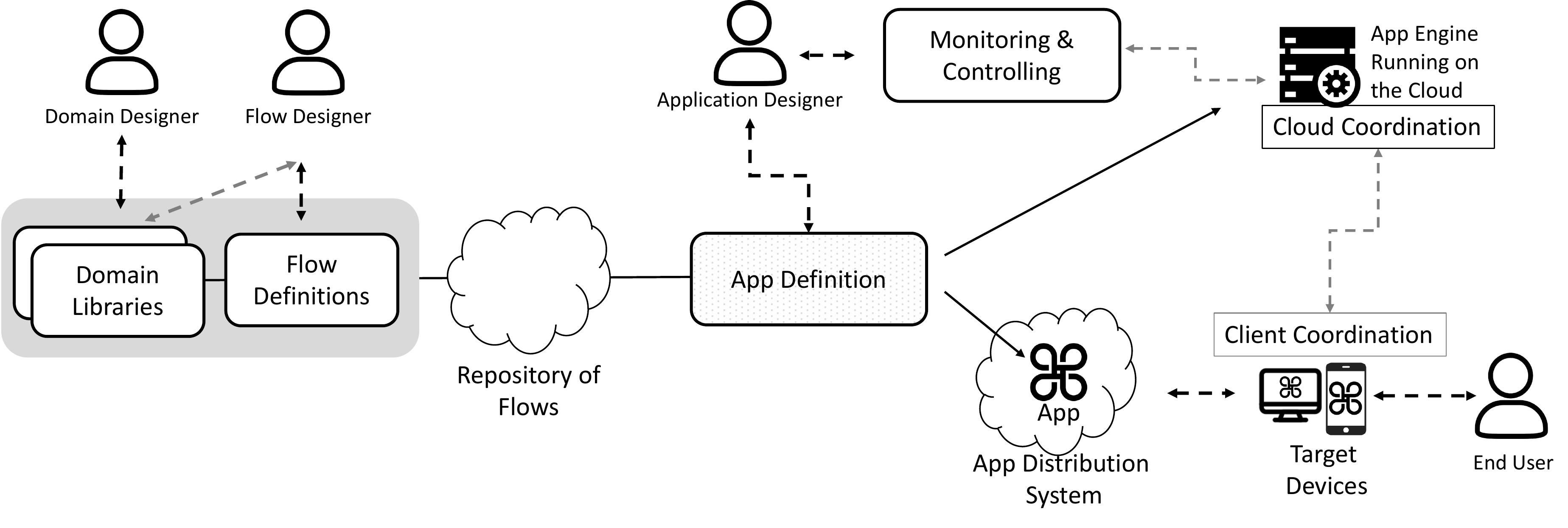}
	\caption{General method}
	\label{fig:fullChain}
\end{figure*}

\section{System Design}

Figure~\ref{fig:fullChain} presents the general workflow of Flow. We define three different development roles based on workflow components: \emph{Domain Designer}, \emph{Flow Designer}, and \emph{Application Designer}. 
The domain designer creates small pieces of basic functionalities, and organize them by domains, and Flow Designers utilise the domains libraries to compose flows that specify the business logic of apps. The App Designer is responsible for composing the final functionality of the app, by selecting the flows, as well as deciding the target platform and distribution details for the app. Each role has different technical requirements, from some technical requirements needed for the Domain Designer to no technical requirements needed for the Application Designer. \emph{Domain Libraries} and \emph{Repository of Flows} are both reusable across apps, which means that non-technical users does not need to start modeling an application by creating a Domain Libraries, but they can directly model an app by using existing flows.

The details and user interfaces of the app modeling are presented in Subsection \ref{subsec:appModeling}. Subsection \ref{subsec:runtime} described the \emph{Coordination Mechanism} that facilitate app distribution and management. Finally, Subsection \ref{subsec:advantages} summarized some consequences and advantages of this system.

\subsection{Application Modeling} 
\label{subsec:appModeling}

\begin{figure}[h]
\centering
\subfloat[Domain Tasks]{\includegraphics[width=\linewidth]{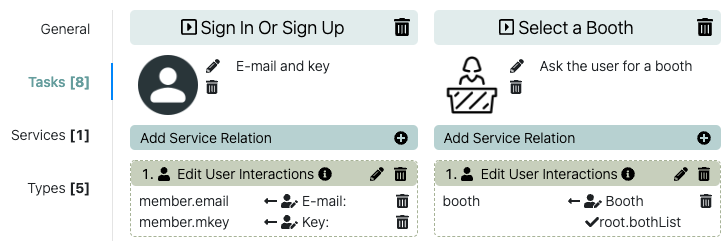}}
\\
\subfloat[Domain Services]{\includegraphics[width=1.1\linewidth]{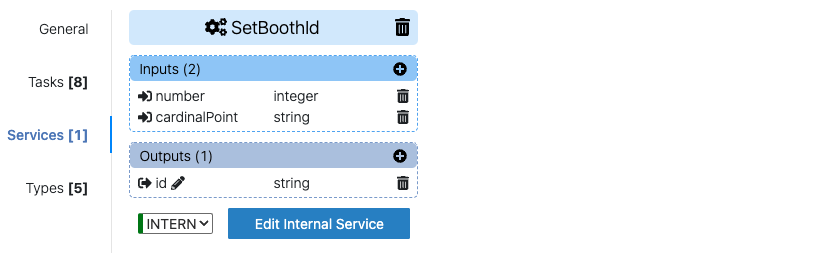}}
\\
\subfloat[Domain Data Types ]{\includegraphics[width=\linewidth]{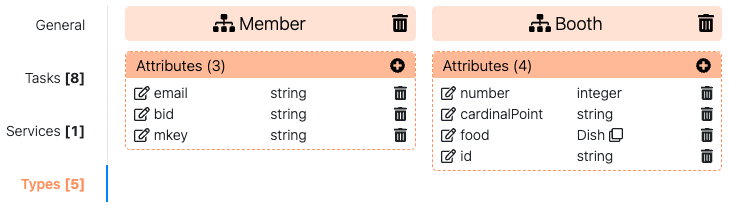}}
\caption{Components in a Domain library}
\label{fig:domain}
\end{figure}

Domain designers create small pieces of reusable logic and group them in \emph{Domains}. A domain consists of a set of data \emph{Types}, \emph{Services} that manage the information and that can be \textit{internal} (defined in the domain) or \textit{external} (provided by third parties), and \emph{Tasks} that are composed for \textit{Service Relations} to call, and \textit{User Iterations}. There are two types of user iterations: \emph{PROMPT} that takes the input of the attribute of the end users and \emph{DISPLAY} that retrieves the attribute value from the app database. Figure~\ref{fig:domain} shows an example of tasks, services, and types in domains that allow end users to sign up for personal accounts.

\begin{figure*}
	\centering
	\includegraphics[width=\textwidth]{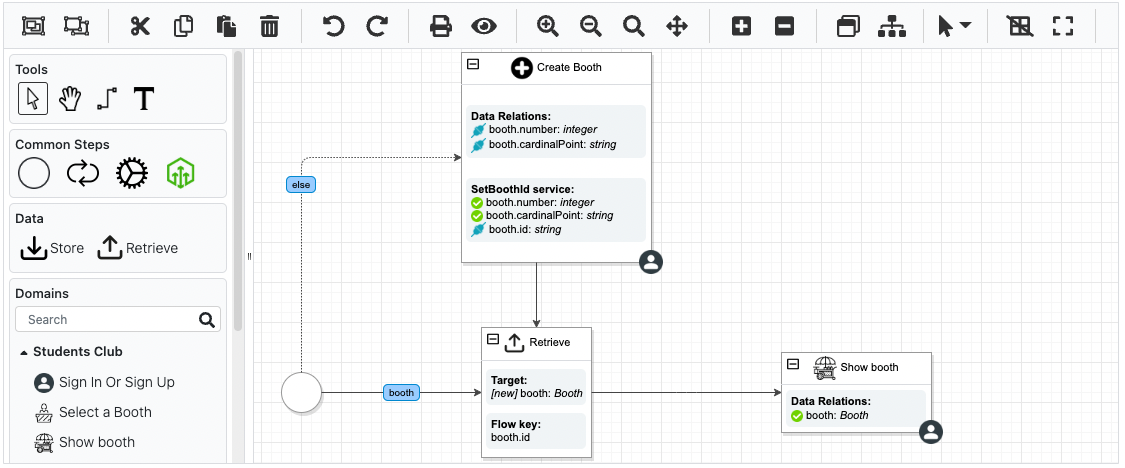}
	\caption{The interface of creating a flow}
	\label{fig:flow-example}
\end{figure*}

Flow designers select one or more domain libraries available in the Repository of Flows, to create reusable high-level behavioral models called \emph{flows}. As can be seen in Figure~\ref{fig:flow-example}, a flow is made up of a set of nodes called \emph{Steps}, and arrows called \emph{Transitions}. There are different types of steps, Common Steps, Data Steps, and Domain Steps, which models the execution of a task defined in a domain. Transitions model the order of execution of steps and can be conditional. For example, specifying a conditional transition can ensure that the step will be executed only if the associated condition is evaluated to be true.

\begin{figure}
	\centering
	\includegraphics[width=\linewidth]{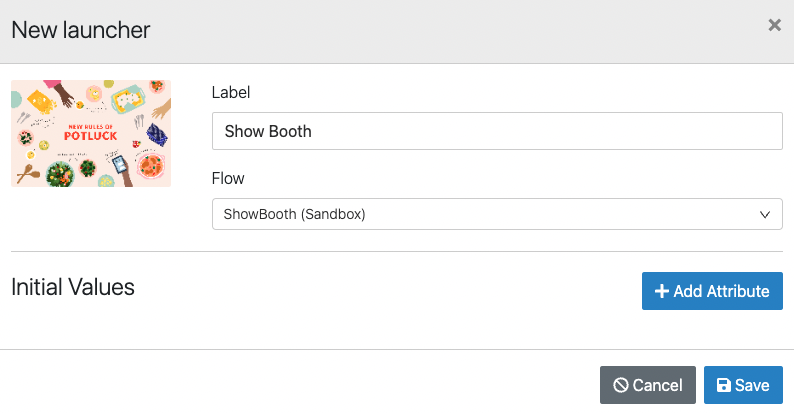}
	\caption{The interface of creating a launcher}
	\label{fig:launcher}
\end{figure}

Finally, to create an application, the Application Designer models the \emph{App Definition}, this involves simple configuration tasks such as setting the application name and logo, binding navigation buttons, and setting static data elements. Specifically, the app designer creates \emph{Launchers} that triggers the execution of a flow in the cloud. As shown in Figure~\ref{fig:launcher}, users select the flow that they want to execute from the Repository of Flows and add meta-information such as label and icon. In addition, they can configure the flow behavior for a particular application by assigning initial values to the attributes contained in the flow.

Once the app is modeled, the Application Designer can decide the target platform where they want to deploy the app. The system offers different methods to distribute apps to end users, including deploying them in a cluster as a web app, publishing them in an app marketplace, or simply downloading them as a local file.

\subsection{Execution and Coordination Mechanism}
\label{subsec:runtime}

As explained above, all flows are executed in the cloud, and the end-user applications are generic and self-adaptable to the requirements of each user iteration. 
This is possible thanks to the Coordination Mechanism, based on the author's prior work (\cite{perez2020abstract,perez2021decoupling}), which provides a layer of abstraction for the end-user applications. This layer is only responsible for displaying and gathering data from the user.
Therefore, the application is always attached to the execution engine running in the cloud where the flows are executed, as shown in Figure~\ref{fig:fullChain}.

This mechanism works as follows: When the end user launches a flow through a client, the engine retrieves the flow model specification and starts executing it.
The cloud coordinator keeps executing the flow on the cloud, in the order defined by the transitions, until a user-iteration need is detected. At this moment, the cloud coordinator sends a message to the end-user application, including the information that the client coordinator will need to proceed.

As can be seen in the example of the following Listing, this information is essentially serialized as a JSON dictionary that contains the elements to display to the end-user (displayElements), the elements to gather from the end-user (gatherElements), any constraints that the client coordinator must satisfy before sending a response (constraints), and data objects to display, and default values for elements to gather (values).

This dictionary does not specify to the end-user clients who interact with the user, and does not contain any HTML, CSS, or other user interface code. The client coordinator has fully delegated the responsibility of interacting with the end-user. Generic end-user applications contain the logic to adapt their interfaces to the end-user iteration needed for each case. For instance, if the end-user application is a skill for a voice-based assistant such as Alexa, the application will render voice commands to interact with the end users, whereas if the end-user application is screen-based, this application probably will generate visual elements to the same end. In this way, the iteration model will depend on the characteristics of the target device, and the flow execution can take advantage of the target platform, such as available sensors.

Flow already provides applications for a set of target platforms, but these generic applications can be developed or customized by third parties with technical skills.

\begin{lstlisting}[language=json, mathescape=true]
{"instanceId": 15,
 $\textbf{"displayElements"}$: []}],
 $\textbf{"gatherElements"}$: [
  {"name":"booth_number", 
   "label":"Booth Number:",
   "set":false, "type":"Integer"},
  {"name":"booth_cardinalPoint", 
  "label":"Cardinal Point:",
   "set":false, "type":"string"}]
 $\textbf{"constraints"}$: [
  {"name":"booth_cardinalPoint", 
  "valueFrom": "cpoints"}]] 
 $\textbf{"value"}$: [
  {"cpoints": ["North","South",
               "East","West"]}]} 
\end{lstlisting}

Once the user iteration is finished and the user has satisfied these conditions using the user interface, the client coordinator is able to send a message back to the cloud coordinator about the result of the interaction. As can be seen in the example of the following Listing, the response is basically another JSON dictionary containing the requested information.

The cloud coordinator receives the information, and now it can continue the execution of the flow in the cloud, for that it determines the next step or finalized if no more steps to execute.

\begin{lstlisting}[language=json, mathescape=true]
{"instanceId": 15,
 $\textbf{"response"}$: [
  {"booth_number":1},
  {"booth_cardinalPoint":"North"}}
\end{lstlisting}

\subsection{Advantages and Consequences}
\label{subsec:advantages}

As a consequence of this mechanism, end-user clients are not aware of the overall execution of the flows happening in the cloud, they only knows the current activity, what data are provided by the cloud coordinator, and what data is needed by the cloud coordinator. 

One of the advantages of our approach is that it is possible to see the results directly on the final application from the first moment, and therefore they can model the app incrementally. But additionally, as can be seen in Figure~\ref{fig:fullChain}, they can also \textit{Monitor and Control} what is being executed in the cloud. 
This is useful during the design because we can observe all the data that have been interchanged, or users iterations during the execution of the flow. But this is also useful in production time, since we can observe how end-users use the application.

Another advantage of our approach is that the application designer can modify the application definition at anytime, and all clients are always automatically updated. 
This is useful not only for non-technical users, but also for the more experienced ones that do not need to maintain the base code for each target-platform if the needs change.

\section{User Evaluation}
In the user evaluation, we focus on the primary design objective of Flow: supporting app development for non-technical people. Therefore, an observational experiment was performed where participants learned how to use Flow and developed a new flow based on instructions. The recruitment, the study setup, and results are presented below.

\subsection{Participants}
We recruited 10 participants from two undergraduate level information science courses for non-computer science major students on a voluntary base. The sample population was chosen as the enrollment in the courses indicates that students have no more than basic programming knowledge but are also interested in learning more. Therefore, even though the population characteristics may not be the same as those of real-world Flow users, the recruited participants are valid for evaluating the user experience. Participants consist of 1 freshman, 4 sophomore, and 5 junior students. All of them are in the 18-24 age group and 6 of them were female. Six participants self-identified as White and four of them Asian. Each participant was compensated \$15 for completing the study. 

\subsection{Study Setup}
Observational interviews based on the think-aloud protocol (\cite{Wright1991use}) were conducted to evaluate the usability of Flow. Participants were asked to perform realistic tasks and verbalize their actions, questions, and relevant thoughts during the process. The study consists of four parts: pre-task survey, pre-task training, task completion, and post-task interview. Data were captured through audio recordings, computer screen recordings, and electronic questionnaires. The study has been reviewed and approved by the Institutional Review Board in the authors' university.

The pre-task survey collected participants' demographic information and previous experience with programming. 
During the pre-task training, we used a realistic app development scenario to familiarize participants with the system. The following scenario descriptions were presented to participants:

\emph{One university student club wants to host a potluck event to introduce the students to cuisines of different cultures and you’ll be creating an application for this event. Below is the requirements for the application:
\begin{itemize}
\setlength\itemsep{0em}
    \item There would be a Sign-Up list for club members to pledge to donate food to the event. They need to enter their name and email address.
    \item There are different booths and the members need to bring food to a specific booth and add descriptions about the food.
    \item Event participants can use the app to access information about the food.
    \item Event participants can use the app to leave ratings for the food. They also need to enter their name and email address.
\end{itemize}
}

This potluck app has been constructed in Flow beforehand. The application consists of 4 launchers: sign up for the event (Sign Up), select the booth to bring food to (Select Booth), check food descriptions (Show Info) and leave a rating for food (Review). A researcher would walk the participants through the development process and answer questions about how to use the system. Once the participants have no more questions, they were asked to complete two tasks:
\begin{description}
\setlength\itemsep{0em}
\item[Task 1] Create a new launcher ``Survey'' using the same flow as the Review launcher. Assign initial values to at least one attribute.
\item[Task 2] Create a new launcher ``Welcome'' by creating your own flow. The launcher can display a static text message and an image to all users of the app.
\end{description}

The first task served as a warm-up exercise to familiarize participants with the interface. The second task urged participants to go through the entire app development process under minimal circumstances: creating a data type, designing a domain, designing a flow, and creating a new launcher for the app. A researcher would be present to answer any questions and assist participants if they could not proceed. 

The post-task interview used the System Usability Score (SUS) questionnaire (\cite{bangor2008empirical}) to measure the usability of Flow. The questionnaire consists of 10 statements that participants need to rate on a Liker scale from  1 (Strongly disagree) to 5 (Strongly agree). To calculate the final score, we subtract one from the user responses for statements 1-5, and subtract the user responses from five for statements 6-10. The converted responses will be added up and multiplied by 2.5. Therefore, the final SUS value will range from 0-100 with each statement 0-10. In addition, we conducted a semi-structured interview to understand participants' feedback. The interview was guided by three main questions: 
\begin{itemize}
\setlength\itemsep{0em}
    \item What was the most difficult thing to understand or do during the task?
    \item What aspects would you change for this system?
    \item Would you use the system to create any applications in real life? 
\end{itemize}

\subsection{Results}
Table~\ref{tab:summary} summarizes the results of the user evaluation. Participants' background in programming or app development were described with the number of years in their primary programming languages. It should be noticed that the number does not reflect their experience of using a programming language for work but rather when they learned it in a university course. The average task completion time was 21 minutes, excluding training and post-task interview and including questions during the task. The average SUS is 61.8 with a highest of 77.5 and a lowest of 32.5. The average score for each criterion is shown in Table~\ref{tab:sus}. The score indicates the level of agreement for each statement and statements 6-10 were reversely coded. 

Through qualitative analysis of audio recordings and screen recordings, we extracted and counted the number of unique critical incidents during evaluations: procedure issues and interface issues. Procedure issues were identified when participants experienced difficulty in identifying the objectives they need to achieve or the procedures they need to execute. Interface issues were identified when participants experienced difficulty in locating specific elements in the interface or experienced unexpected UI behaviors. Negative and positive comments regarding their experience with the system were identified from post-task interviews. The implications of each measurement are detailed below.

\begin{table*}[ht]
\centering
\caption{Participants' background in programming and the summary of user evaluation measurements.}
\label{tab:summary}
\begin{tabular}{lcccccclc}
\hline
\multicolumn{1}{c}{No.} &
  Background &
  \begin{tabular}[c]{@{}c@{}}Completion\\ Time (mins)\end{tabular} &
  SUS &
  \begin{tabular}[c]{@{}c@{}}Procedure\\ Issues\end{tabular} &
  \begin{tabular}[c]{@{}c@{}}Interface\\ Issues\end{tabular} &
  \multicolumn{2}{c}{\begin{tabular}[c]{@{}c@{}}Positive\\ Comments\end{tabular}} &
  \begin{tabular}[c]{@{}c@{}}Negative\\ Comments\end{tabular} \\ \hline
P1  & \textless 1-year Matlab & 27 & 72.5 & 5 & 5 & \multicolumn{2}{c}{2} & 3 \\
P2  & 4-year HTML             & 22 & 57.5 & 1 & 4 & \multicolumn{2}{c}{3} & 1 \\
P3  & 1-year Java             & 12 & 70.0   & 0 & 1 & \multicolumn{2}{c}{3} & 4 \\
P4  & 1-year Java             & 21 & 75.0  & 3 & 3 & \multicolumn{2}{c}{3} & 2 \\
P5  & 1-year SQL              & 22 & 32.5 & 9 & 2 & \multicolumn{2}{c}{0} & 0 \\
P6  & \textless 1-year C++    & 22 & 72.5 & 3 & 1 & \multicolumn{2}{c}{1} & 2 \\
P7  & 3-year Python           & 12 & 72.5 & 2 & 5 & \multicolumn{2}{c}{1} & 2 \\
P8  & \textless 1-year SQL    & 28 & 42.5 & 8 & 2 & \multicolumn{2}{c}{2} & 3 \\
P9  & 2-year Java             & 21 & 77.5 & 2 & 5 & \multicolumn{2}{c}{2} & 2 \\
P10 & 1-year Python           & 24 & 45.0   & 6 & 2 & \multicolumn{2}{c}{0} & 5 \\ \hline
\end{tabular}
\end{table*}

\begin{table*}[ht]
\centering
\caption{Average scores of individual statements in the SUS scale. 6-10 were reversely coded (*).}
\label{tab:sus}
\begin{tabular}{llc}
\hline
\multicolumn{1}{c}{} & \multicolumn{1}{c}{Statements}                                                 & \multicolumn{1}{c}{Average Scores} \\ \hline
1                    & I think I would use this system frequently                                   & 6.3                       \\
2                    & I thought the system was easy to use                                         & 6.5                       \\
3                    & I found the various functions in this system were well integrated            & 8.8                       \\
4                    & I would imagine that most people would learn to use this system very quickly & 6.8                       \\
5                    & I felt very confident using the system                                       & 3.3                       \\
6*                    & I found the system unnecessarily complex                                     & 5.8                       \\
7* & I think that I would need the support of a technical person to be able to use this   system & 5.0 \\
8*                   & I thought there was too much inconsistency in this system                    & 7.5                       \\
9*                   & I found the system very cumbersome to use                                    & 6.8                       \\
10*                   & I needed to learn a lot of things before I could get going with this system  & 5.3                       \\ \hline
\end{tabular}
\end{table*}

\textbf{System Usability Score.} According to the adjective rating scale proposed by \textcite{bangor2009determining}. The average SUS of 61.8 falls between the middle of ``OK'' and ``Good''. Three participants (P5, P8, and P10) rated the usability as below ``OK''. Their background in programming is ``$\leq$ 1-year SQL'' or ``1-year Python''. By contrast, most participants who gave a rating greater than 70 had experience with Java or C++ (P3, P4, P6, P9). This might suggest that the experience with certain programming languages helped users adapt to the system. The score of individual statements suggested that participants may need to learn a lot of things (10) and require the support of a technical person (7) for completing the tasks. Therefore, they were not very confident in using the system (5).

\textbf{Procedure Issues.} The number of procedure issues reflected participants' understanding of the app development process and related concepts in Flow. For example, the main issues are ``the functionality of initial values'' (n = 7), and ``the functionality of the data relations'' (n = 7). Participants seemed confused about the use of initial values as they would assign values to random attributes regardless of the flow they are aiming to launch or even thought that they are leaving a comment in the app as app users. Another relevant issue is ``the functionality of labels'' (n = 4), where participants would confuse the metadata of domain (descriptions and labels) with initial values displayed to app users. Additionally, participants may have difficulty associating data relations with application functionalities. For example, after enabling the DISPLAY data relation in the domain, P2 still tried to ``add an operation that will display the message" in the flow. These issues indicate that the workflow and terminology of Flow may not be intuitive to users.

\begin{figure}[h]
\centering
\subfloat[The Define button and the Edit button seem similar]{\includegraphics[width=\linewidth]{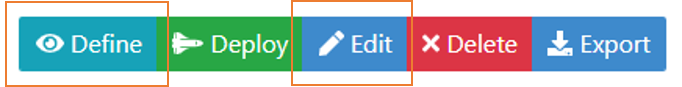}}
\\
\subfloat[The two plus icons seem similar ]{\includegraphics[width=\linewidth]{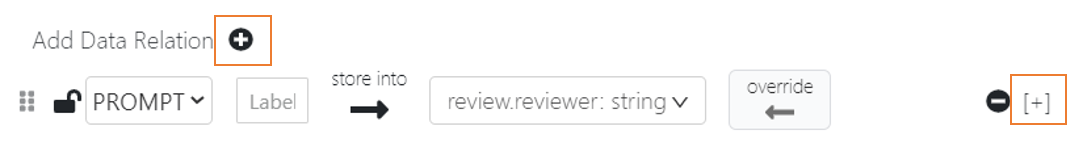}}
\caption{Examples of ambiguous objects}
\label{fig:ambiguous}
\end{figure}

\begin{figure}[h]
\centering
\subfloat[Assign initial value to a single attribute]{\includegraphics[width=\linewidth]{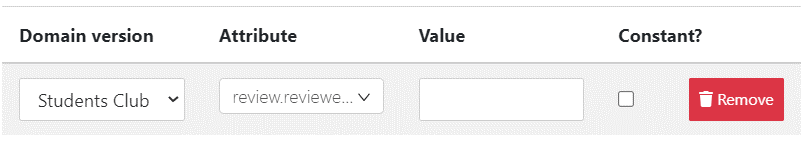}}
\\
\subfloat[Assign initial values to a group of attributes ]{\includegraphics[width=\linewidth]{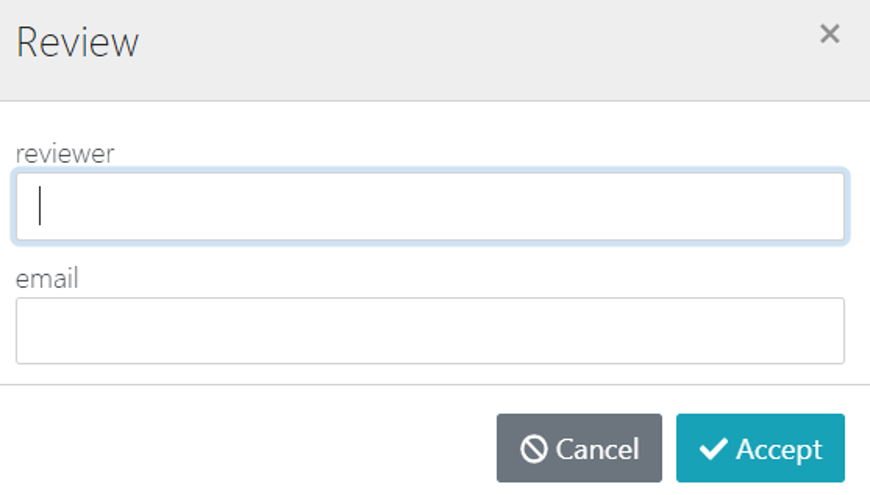}}
\caption{Two approaches to assign initial values}
\label{fig:dropdown}
\end{figure}

\textbf{Interface Issues.} The user evaluations also suggested room for further improvements in the UI design. One major issue is ``ambiguous objects'' (n = 8). Some buttons with different functionalities may seem similar to participants and they could find it difficult to decide which one they should click, see Figure~\ref{fig:ambiguous} for examples. In addition, there are two potential approaches to assign initial values (see Figure~\ref{fig:dropdown}), yet it turned out that the existence of alternatives could confuse the participants. Another issue is the ``lack of signifier'' (n = 7). Participants would ask questions such as ``Did I click the application (tab)? (P1)'' or ``is this saved or not? (P7)''. More visual cues may be needed to signify the action status and the object affordances in the interface. Interface issues suggest that the current UI design may not efficiently help users learn how to use Flow.

\textbf{Positive Comments.} The major positive comments are ``the system is easy/straightforward to use'' (n = 7), ``the system is helpful for non-programmers'' (n = 4), and ``app development through Flow is easier than coding'' (n = 3). Specifically, participants agreed that the system is ``pretty easy to use once you know how to do it (P3)'', but it is also necessary to ``do it a couple of times (P2)'' and ``learn where things are located (P1)''. Some commented that this is ``good for people that have no coding experience (P2)'' and ``it seems to streamline the process that otherwise would take a long time... I cannot make something like this through code (P4)''. The initial results suggest that Flow has simplified the app development process and made it more accessible to people with little or no technical background.

\textbf{Negative Comments.}  The major negative comments are ``the system requires learning time'' (n = 8) and ``unnecessary steps in the workflow'' (n = 7). Specifically, most participants have mentioned that ``things were confusing at first (P8)'' or ``understanding this (domain) was hard, I wasn't sure what to do at first (P6)''. This did not conflict with the positive comments. In combination, they suggest that there could be a steep learning curve at the initial phase but the system became straightforward after this phase. The second theme partially explained the learning curve: ``Going back between these (data types and domains) are cumbersome (P6)'' and ``the most difficult thing was understanding the order in which I had to do everything (P3)''. As a result, participants have suggested providing a ``step-by-step tutorial (P4)'' or ``highlighting the sequence (of doing tasks) (P8\&P9)''. Negative comments indicate that more efforts are needed to reduce the amount of workload for users. 

In addition, 8 participants expressed that they became more interested in learning app development after using Flow and that they are likely to use Flow in real life. Three statements were used to capture their satisfaction with the system and participants showed a moderate level of average agreement: ``I would recommend this system'' (3.8/5), ``I am satisfied with this system'' (3.9/5), and ``I enjoyed using this system to create an application'' (4.1/5).

\section{Discussion}
In this section, key findings of the Flow framework and user evaluations are discussed with regard to prior research and main contributions of this paper. The definition of behavior as models or high-level abstractions has been frequently studied in the literature, and many of them used DSLs or graphical languages (\cite{van2005yawl,bocciarelli2011bpmn,braun2014classification}). By contrast, our approach not only allows the no-code definition of business logic but also provides a coordination mechanism to separate execution and the client-side representation. In terms of cross-platform app development, existing tools (\cite{bosnic2016development,boduch2017react,hermes2015xamarin}) can support developing mobile apps that run on multiple platforms. However, the levels of abstraction differ significantly and they require relatively high technical skills. Therefore, our work enhanced cross-platform app development by customizing it for non-technical users.

Some studies have targeted at supporting users with less technical background. For example, the App Inventor (\cite{patton2019app}) uses a What You See Is What You Get (WYSIWYG) interface and allows users to create Android apps. However, this tools aims at promoting programming education for newcomers rather than developing business apps. The MAML framework proposed by \textcite{rieger2018evaluating} aims to balance technical complexity and graphical simplification when describing cross-platform business applications. Despite the positive results in its usability evaluation, the authors acknowledged that some expertise is needed in data model inference. Moreover, none of the above works integrates cloud-based execution and facilitates the maintenance and monitoring of apps.

In terms of the usability, the average SUS of our framework is comparable to that of MAML \cite{rieger2018evaluating}, although the qualifications of participants are different. Our participants do not have experience in app development or process modelling. One participant has knowledge in web development with HTML, and the rest of them only learned basic programming from introductory courses. The positive comments showed that Flow is easy for these users to operate and successfully lowers the threshold of app development. However, negative comments also suggest some problems with the current design. The terminology used in Flow and the lack of signifiers make the initial learning curve rather steep. Consequently, participants felt that they needed considerable time and help to familiarize themselves with the system. This could be solved by optimizing the workflow and the UI design, as illustrated in procedure issues and interface issues. Specifically, the workflow could be further simplified to reduce the number of steps needed, the technical terms could be replaced or explained with hover-over text, and visual cues could be integrated in the UI to guide the users. The results demonstrated that the Flow is friendly to at least users with little technical background, yet users with some technical knowledge may experience fewer problems when learning Flow.

There are also limitations in this paper and studies are ongoing to address them. First, our participants are undergraduate students with basic programming knowledge. Their experiences and attitudes could be different from the non-technical who need to develop apps for business purposes. In addition, this user evaluation did not cover all aspects of Flow. The compatibility with more complex platform interactions, deployments, and the maintenance of apps should be evaluated in future research. 

\section{Conclusion}

In this paper we present and evaluate a no-code solution that enables cross-platform app development for non-technical users. Compared with frameworks proposed in prior research, Flow does not generate code for DSL models specified by users. Instead, the models will be executed directly on the cloud. There are several advantages in our approach. The app execution is based on the business logic defined by users, which makes it possible to manage a set of clients for different platforms, and any changes made in the logic will be updated simultaneously for all clients. Non-technical users can easily develop apps without learning programming languages or other technical knowledge and also monitor the app execution and manage the maintenance with Flow.

The user evaluation demonstrates that the current prototype has reached a good usability and users felt positive about developing apps using Flow. However, some knowledge in programming languages could help them to learn how to use Flow and the evaluation highlighted design directions for further improving the user experience. There are ambiguous UI objects, and it takes time before users become familiar with the system. Simplification of the workflow and adding signifiers in UI design could reduce the workload, especially in the early stage of learning. In terms of study limitations, a more representative participant sample for real-world users who need business apps can strengthen the validity of the system usability. In addition, we plan to evaluate the full functionality of Flow in the future, as well as the collaboration between three designers in the workflow. 

In conclusion, we believe that this approach makes cross-platform app development more accessible to non-technical users. Although existing usability issues can prevent users from understanding the system in a quick manner, most of them can be fixed through optimization of the UI. Hence, Flow can potentially reduce the cost in app development, and support a larger population to participate in business where existence in app marketplace is playing an important role.

\printbibliography

\end{document}